\documentclass[pra,aps,twocolumn,showpacs,10pt]{revtex4-1}
\usepackage{amsmath,amssymb,graphicx,bm,color,ulem}

\begin{document}

\title{Attosecond helical pulses}

\author{Miguel A. Porras}

\affiliation{Grupo de Sistemas Complejos, ETSIME, Universidad Polit\'ecnica de Madrid, Rios Rosas 21, 28003 Madrid, Spain}

\begin{abstract}
We find a solution of the wave equation in the paraxial approximation that describes the attosecond pulses with spatiotemporal helical structure in the phase and in the intensity recently generated by means of highly nonlinear optical processes driven by visible or infrared femtosecond vortex pulses. Having a simple analytical model for these helical pulses will greatly facilitate the study of their predicted applications, particularly their interaction with matter after their generation. It also follows from our analysis that the topological charge dispersion inherent to helical pulses allows to beat the minimum duration to which a pulsed vortex without charge dispersion is limited.
\end{abstract}

%\pacs{32.80.Wr; 42.65.Tg; 05.45.Yv}

\maketitle

\section{Introduction}

In recent years there have been significant advances in the generation of extreme ultraviolet and x-ray attosecond pulses with orbital angular momentum (OAM) by means of highly nonlinear processes driven by visible or infrared femtosecond pulses carrying also OAM \cite{HERNANDEZ,GARIEPY,HERNANDEZ2,GENEAUX,REGO}. It has been demonstrated that the natural structure of the attosecond pulses that result from the coherent superposition of high harmonics of different frequencies and OAM is a helical spatiotemporal structure in both the phase and the intensity. The general properties of these helices of light, also called ``light springs", and their relevance for applications have been detailed in \cite{PARIENTE}. The description in \cite{PARIENTE} is however qualitative in many aspects, with no analytical expressions or numerically evaluated intensity or phase profiles or their changes during propagation; indeed most of the description refers to a single transversal plane, propagation effects then being absent.

Here we provide a simple analytical expression of helical pulses or light springs satisfying the paraxial wave equation for superbroadband light propagating in free space, and describe their spatiotemporal structure and propagation features. As the word ``spring" suggests something elastic  but these helices of radiation have a fixed pitch, we prefer to refer to them as helical pulses. Although we focus in the attosecond time scale and in the specific conditions that reproduce the structure of the attosecond pulses generated in experiments, the same expression holds at other time scales at visible or infrared carrier wavelengths, and with other conditions determined by the free parameters involved in the analytical expression. In the same way as with other fundamental luminous objects such as Gaussian beams and pulses, Laguerre-Gauss beams, Bessel beams, etc, having a simple analytical expression of these helical pulses will facilitate (e. g., will eliminate the necessity of performing costly high-harmonic generation numerical simulations) theoretical studies of their expected applications such as the excitation of attosecond electron beams carrying OAM \cite{GENEAUX} or transfer of OAM to matter by stimulated Raman scattering \cite{PARIENTE}.

Another important issue is the duration of the individual pulses in the helical structure. As recently found, \cite{PORRAS5,PORRAS6} a pulsed vortex with well-defined topological charge, i. e., without topological charge dispersion, must be longer than a certain minimum value determined by the topological charge. Helical pulses are superpositions of pulsed vortices with carrier frequencies in a frequency comb and with topological charges varying linearly with frequency, and therefore present topological charge dispersion. We show how to manage this dispersion to synthesize attosecond pulse trains, or isolated attosecond pulses, with a certain mean topological charge that are shorter than the minimum duration of a pulsed vortex of the same charge without dispersion. Indeed there is no lower bound to the pulse duration as long as sufficiently high topological charge dispersion is introduced.

\section{Cylindrically symmetric pulsed vortices}

We consider ultrashort, three-dimensional wave packets, $E(x,y,z,t)$, propagating mainly in the positive $z$ direction. Introducing the local time $t'=t-z/c$, where $c$ is the speed of light in vacuum, the wave equation $\Delta E -(1/c^2)\partial^2_t E=0$ reads as  $\Delta E = (2/c)\partial^2_{zt'} E$. The so-called pulsed beam equation \cite{HEYMAN,PORRAS1,BESIERIS}, or paraxial wave equation for ultrashort wave packets,
\begin{equation}\label{PBE}
\Delta_\perp E = \frac{2}{c}\frac{\partial^2 E}{\partial z\partial t'}\, ,
\end{equation}
($\Delta_\perp=\partial^2_x+\partial^2_y$ is the transverse Laplace operator) is obtained by neglecting $\partial_z E$ compared to $(1/c)\partial_{t'} E$. This approximation is valid as long as the characteristic axial length of variation of $E$ due to diffraction is much larger than the characteristic axial length of variation of the wave form, e. g., diffraction changes are negligible in a single axial ondulation \cite{PORRAS1}. Also, writing $E=Ae^{-i\omega_0 t'}$, where $\omega_0$ is a carrier frequency, Eq. (\ref{PBE}) would yield the envelope equation introduced in \cite{BRABEC} particularized to free space. The general solution of Eq. (\ref{PBE}) can be expressed as the superposition of pulsed beams
\begin{equation}\label{AZ}
E=\sum_{j} \tilde a_j E_{l_j}(r,z,t')e^{il_j\phi}\,,
\end{equation}
of different integer topological charges $l_j$, where $\tilde a_j$ are arbitrary complex weights, and $(r,\phi,z)$ are cylindrical coordinates. The cylindrically symmetric pulsed vortices $E_le^{il\phi}$ satisfy
\begin{equation}\label{PBEL}
\frac{\partial^2 E_l}{\partial r^2} + \frac{1}{r}\frac{\partial E_l}{\partial r} -\frac{l^2}{r^2}E_l = \frac{2}{c}\frac{\partial^2 E_l}{\partial z \partial t'}\,.
\end{equation}
Writing them as superpositions of monochromatic vortex beams
\begin{equation}\label{AN}
E_l(r,z,t')e^{il\phi}=\frac{1}{\pi}\int_0^{\infty} \hat E_{l,\omega}(r,z)e^{-i\omega t'} d\omega \, e^{il\phi}\, ,
\end{equation}
of angular frequencies $\omega$, the monochromatic constituents, $\hat E_{l,\omega}(r,z)$, must satisfy the paraxial wave equation
\begin{equation}\label{PWE}
\frac{\partial^2 \hat E_{l,\omega}}{\partial r^2} + \frac{1}{r}\frac{\partial \hat E_{l,\omega}}{\partial r} -\frac{l^2}{r^2}\hat E_{l,\omega} +2i\frac{\omega}{c}\frac{\partial\hat E_{l,\omega}}{\partial z}=0 \, .
\end{equation}
Particular solutions to Eq. (\ref{PWE}) are Laguerre-Gauss beams of zero radial order, given by,
\begin{equation}\label{LG}
\hat E_{l,\omega}(r,z) = \hat b_\omega \frac{e^{-i(|l|+1)\psi(z)}}{\sqrt{1+\left(\frac{z}{z_R}\right)^2}}\left(\frac{\sqrt{2}r}{s_\omega(z)}\right)^{|l|}e^{\frac{i\omega r^2}{2cq(z)}} \, ,
\end{equation}
where $q(z)=z-iz_R$ is the complex beam parameter, $\psi(z)=\tan^{-1}(z/z_R)$ is Gouy's phase, and $z_R$ is the Rayleigh distance, which will be assumed to be independent of the frequency, i. e., we adopt the so-called isodiffracting model \cite{PORRAS1,PORRAS2,FENG,PORRAS3,PORRAS4}. The complex beam parameter is often expressed as
\begin{equation}\label{Q}
\frac{1}{q(z)} = \frac{1}{R(z)} + i\frac{2c}{\omega s_\omega^2(z)}
\end{equation}
where $R(z)=z+ z_R^2/z$ is the radius of curvature of the wave fronts, $s_\omega(z)=s_\omega\sqrt{1+(z/z_R)^2}$ is the Gaussian width of the fundamental ($l=0$) Gaussian beam, and $s_\omega=\sqrt{2z_R c/\omega}$ is the waist width located at $z=0$.

It has recently been demonstrated \cite{PORRAS5} that the pulsed vortex of topological charge $l$ in Eq. (\ref{AN}), obtained as superpositions of Laguerre-Gauss beams (\ref{LG}) of different frequencies with adequate weights, with a prescribed pulse shape
\begin{equation}
P(t)=A(t)e^{-i\omega_0 t} = \frac{1}{\pi}\int_0^\infty \hat P_{\omega} e^{-i\omega t} d\omega
\end{equation}
at the caustic surface or revolution hyperboloid $r_p(z)=\sqrt{|l|/2}\, s_{\omega_0}(z)$ of maximum pulse energy, or bright caustic surface surrounding the vortex, is given by the expression \cite{PORRAS5}
\begin{eqnarray}\label{FIXED}
E_l(r, z, t') e^{il\phi} &=& \frac{e^{-i(|l|+1)\psi(z)}e^{il\phi}}{\sqrt{1+\left(\frac{z}{z_R}\right)^2}}\left[\frac{r}{r_p(z)}\right]^{|l|}
\!\!A\left(t_c\right)e^{-i\omega_0 t_c} \nonumber \\
&=& \frac{e^{-i(|l|+1)\psi(z)}e^{il\phi}}{\sqrt{1+\left(\frac{z}{z_R}\right)^2}}\left[\frac{r}{r_p(z)}\right]^{|l|}
\!\!P\left(t_c\right)
\end{eqnarray}
where $t_c=t'- r^2/2cq(z)+i|l|/2\omega_0$ is as space-dependent, complex time. It has also been demonstrated \cite{PORRAS5} that such a pulsed vortex with a well-defined topological charge $l$ and pulse shape $P(t)=A(t)e^{-i\omega_0 t}$ at the bright caustic surface exists only if $\Delta\omega_A^2 < 4\omega_0^2/|l|$, where
\begin{equation}
\Delta\omega_A = 2\left[\frac{\int_0^\infty |\hat P_\omega|^2(\omega-\omega_0)^2 d\omega}{\int_0^\infty |\hat P_\omega|^2 d\omega}\right]^{1/2}
\end{equation}
is the Gaussian-equivalent half bandwidth (yielding the $1/e^2$ decay half width of $|P_\omega|^2$ for a Gaussian-like spectrum) of the pulse spectrum, and the carrier frequency is defined by
\begin{equation}
\omega_0 = \frac{\int_0^\infty |\hat P_\omega|^2\omega d\omega}{\int_0^\infty |\hat P_\omega|^2 d\omega}\,.
\end{equation}
The above upper bound for the pulse bandwidth implies that an arbitrarily short pulse cannot carry a vortex of the topological charge $l$, but there is lower bound to its duration \cite{PORRAS5,PORRAS6}. As shown below, the dispersion or uncertainty in the topological charge inherent to  the helical pulses will allow to beat these upper and lower bounds of the spectral bandwidth and pulse duration.

\begin{figure}[b]
\begin{center}
  \includegraphics[width=7cm]{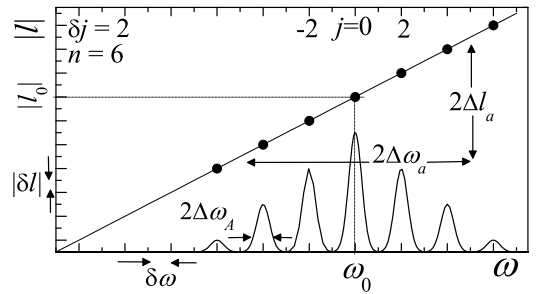}
\end{center}
\caption{\label{Fig1} Scheme illustrating the meaning of symbols used in the text.}
\end{figure}

\section{Helical pulses}

\begin{figure*}[t]
\begin{center}
  \includegraphics*[width=5.5cm]{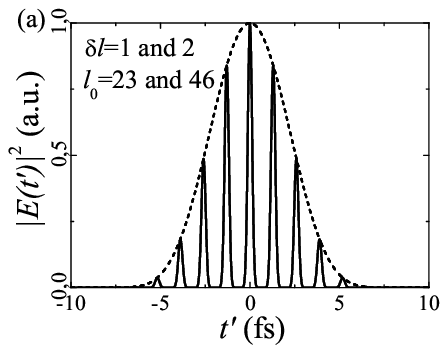}\includegraphics*[width=5.5cm]{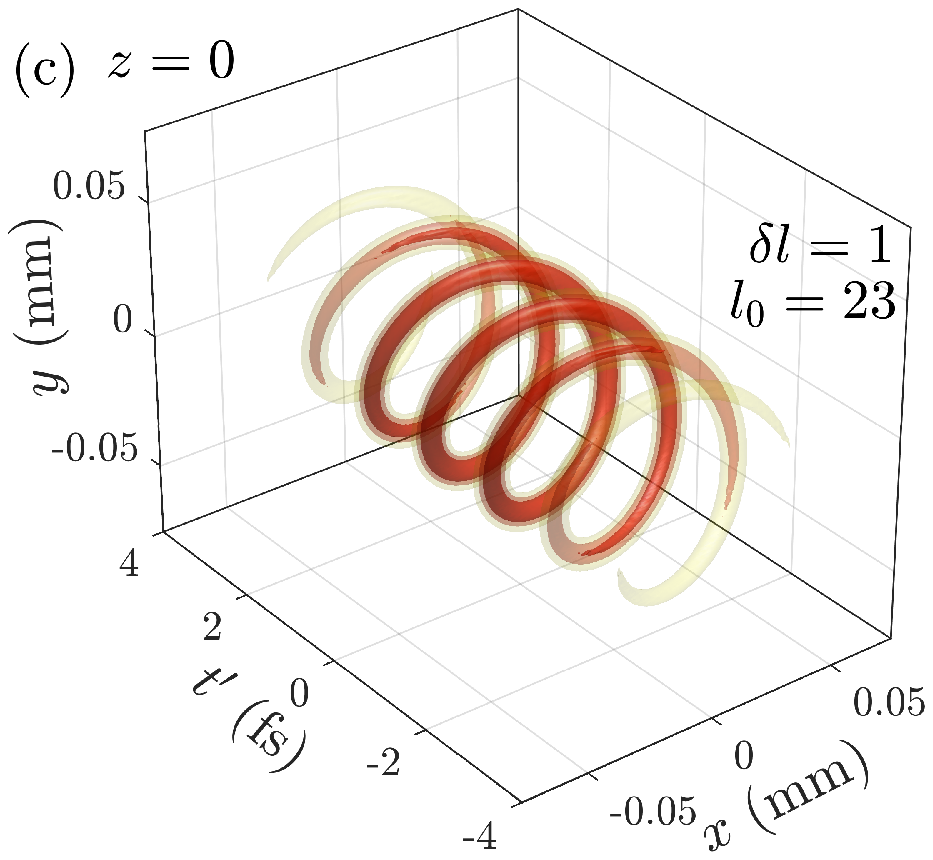}\includegraphics*[width=5.5cm]{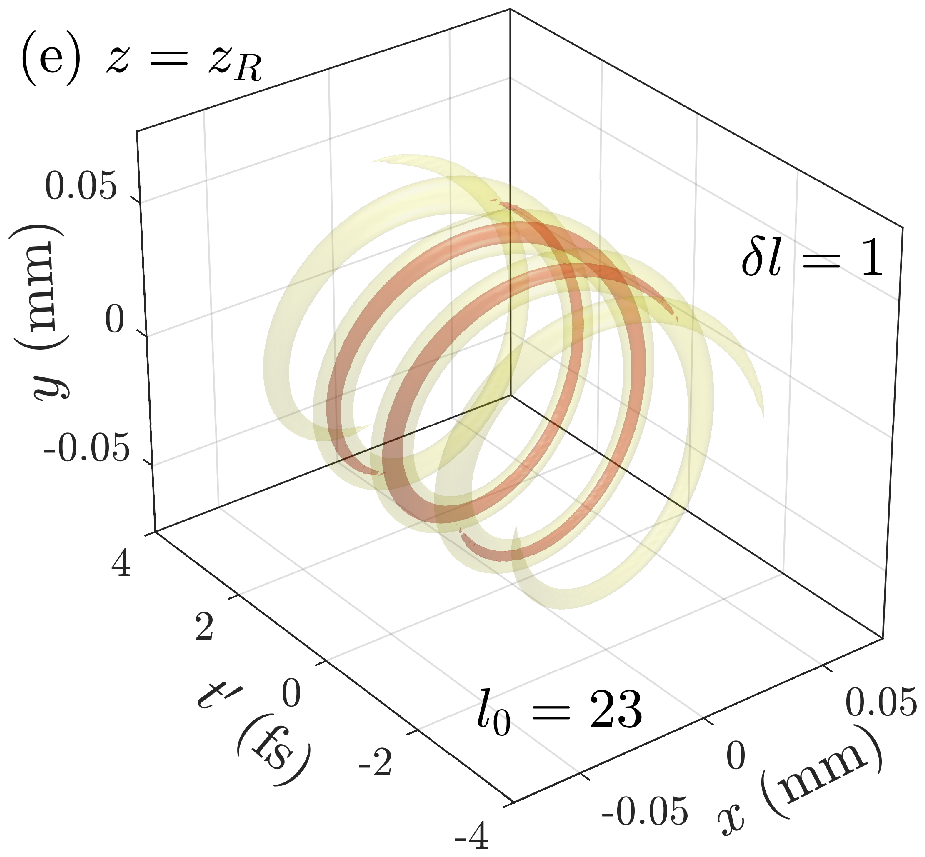}
  \includegraphics*[width=5.5cm]{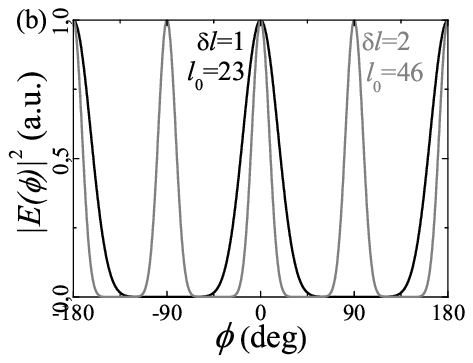}\includegraphics*[width=5.5cm]{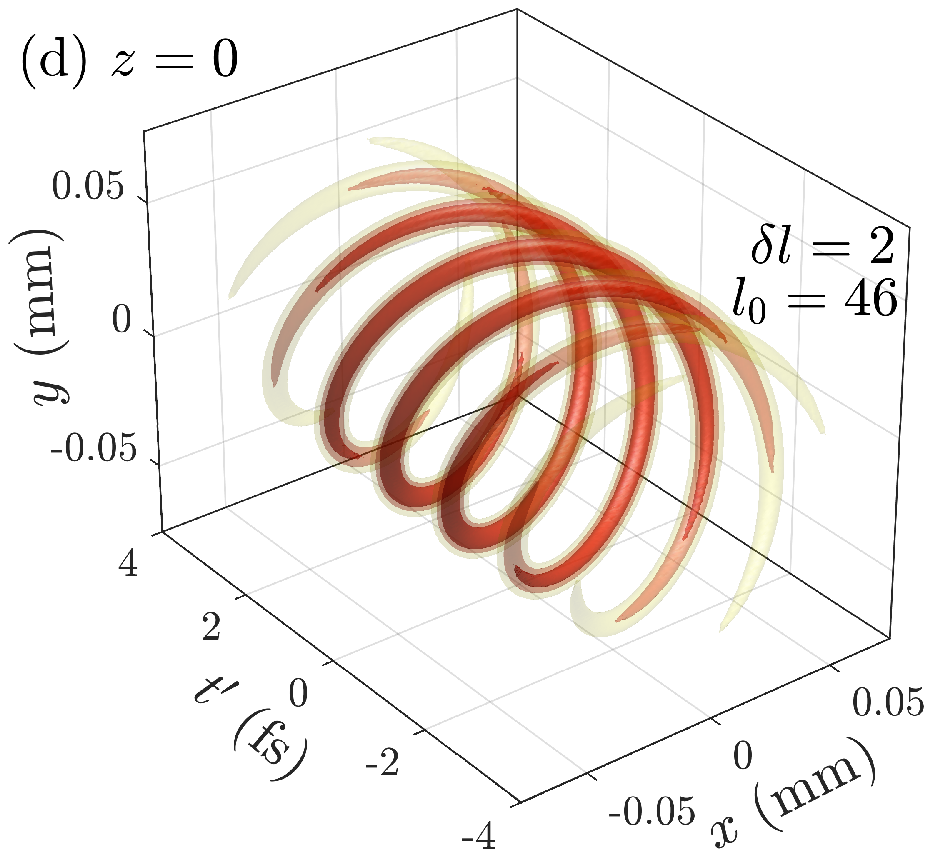}\includegraphics*[width=5.5cm]{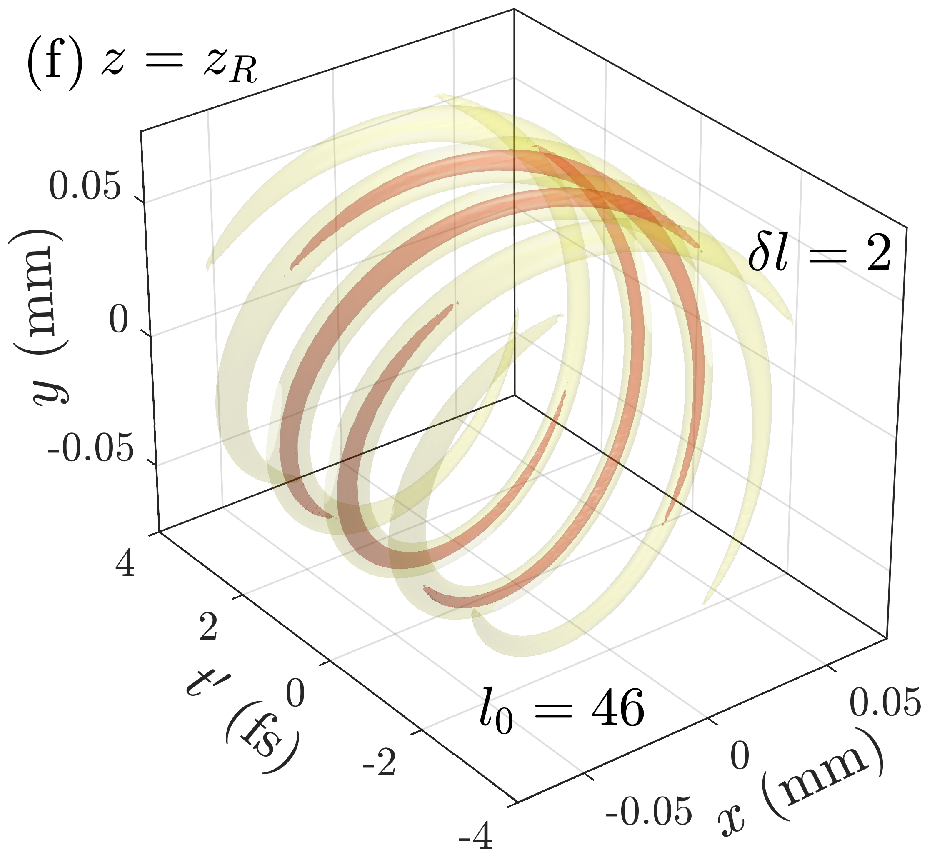}
\end{center}
\caption{\label{Fig2} Spatiotemporal structure of two helical pulses. For both $\delta\omega=2.4166$ rad/fs$^{-1}$ ($780$ nm), $\omega_0=23\delta\omega=55.582$ rad/fs$^{-1}$, $A(t)e^{-i\omega_0 t}=\mbox{sinc}^2(t/T_A)e^{-(i\omega_0t +i\pi/2)}$ with $T_A=8$ fs (two-cycle pulse at $780$ nm), $a(t)=\cos^n(\delta \omega t)$ with $n=6$, and $z_R=10$ mm. In the first case $\delta l=1$ and $l_0=23\delta l=23$, and in the second $\delta l=2$ and $l_0=23\delta l= 46$. (a) Intensity of the attosecond pulses at $\phi=0$ and $z=0$ (black curve) and its femtosecond envelope $|A|^2$ (dashed curve) in both cases; (b) Intensity at $z=0$ and $t'=0$ as a function of the azimuthal angle showing two pulses for $\delta l=1$ and four pulses for $\delta l =2$. (c-f) Spatiotemporal structure of the intensity with 2 intertwined helices for $\delta l=1$ and with 4 intertwined helices for $\delta l=2$, at the waist and at $z_R$. The three surfaces have intensities 0.2, 0.4 and 0.6 times the peak intensity.}
\end{figure*}

We consider now the superposition in Eq. (\ref{AZ}) of the pulsed vortices in Eq. (\ref{FIXED}) of different topological charges and carrier frequencies
\begin{equation}\label{LJ}
l_j=l_0+j\delta l, \quad \omega_j=\omega_0 +j\delta \omega\, ,
\end{equation}
where $j$ are integers about $0$, and $l_0$ and $\delta l$ are integers. For clarity, the meaning of all relevant quantities defined throughout this paper is illustrated in Fig. \ref{Fig1}. Also, the symbols $\Delta \omega$ or $\Delta t$, with or without subindexes, are reserved to Gaussian-equivalent half-widths, and other symbols are used for other measures of width of a function.

We focus on the experimentally relevant situation in which the points ($\omega_j,l_j)$ in the $\omega$-$l$ plane lie in a straight line crossing the origin, implying that $l_j$, $l_0$ and $\delta l$ are either all positive or all negative, and
\begin{equation}\label{COND}
\frac{|l_j|}{\omega_j}=\frac{|l_0|}{\omega_0}=\frac{|\delta l|}{\delta\omega}\,.
\end{equation}
This choice reproduces the conditions of high harmonic and attosecond pulse generation \cite{HERNANDEZ,GARIEPY,HERNANDEZ2,GENEAUX} with a fundamental, visible or near infrared, femtosecond, pulsed vortex if we identify $\delta\omega$ and $\delta l$ with the carrier frequency and topological charge of the fundamental pulse, and $\omega_0=m\delta\omega$ and $l_0=m\delta l$ with the carrier frequency and charge of the $m$th harmonic about the middle of the plateau region in the harmonic spectrum. This spectrum is typically of the form of a frequency comb with tines of similar linewidth \cite{HERNANDEZ}. It is then reasonable to choose $A_j(t)\equiv A(t)$ independent of $j$ so that $\Delta t_{A_j}\equiv\Delta t_{A}$ and $\Delta\omega_{A_j}\equiv \Delta\omega_A$, with $\Delta\omega_A < \delta\omega$ for a comb spectrum, as the simplest, physically reasonable model. Since any fundamental pulsed vortex of frequency $\delta \omega$, charge $\delta l$ and envelope $A(t)$ at its bring caustic surface necessarily satisfies $\Delta \omega_A^2 < 4\delta\omega^2/|\delta l|$, use of Eq. (\ref{COND}) leads to $\Delta \omega_A^2 < 4 (\omega_j/|l_j|)\delta\omega < 4\omega_j^2/|l_j|$, i. e., all superposed cylindrically symmetric pulsed vortices can have the pulse shape $A(t)e^{-i\omega_j t}$ at their bright caustic surface of radius $r_{p,\omega_j}(z)=\sqrt{|l_j|/2}\, s_{\omega_j}(z)$. Further, the choice of $z_R$ independent of $j$ ensures that the bright caustic surfaces of all superposed pulsed vortices overlap, at the waist and during the whole propagation, with that of the fundamental infrared pulse, as expected from the nonlinear interactions generating high harmonics and attosecond pulses, and as described, e. g., in \cite{HERNANDEZ,GARIEPY,HERNANDEZ2,GENEAUX}. In fact, the radii $r_{p,\omega_j}(z)=\sqrt{|l_j|/2}\, s_{\omega_j}(z)=\sqrt{2z_Rc|l_j|/2\omega_j}\sqrt{1+(z/z_R)^2}$ are, on account of Eqs. (\ref{COND}), all equal to $r_{p,\omega_0}(z)$ and to $r_{p,\delta\omega}(z)$, and we can simply write
\begin{equation}\label{RP}
r_{p,\omega_j}(z) \equiv r_p(z)=\sqrt{\frac{|l_0|}{2}}s_{\omega_0}(z)\,.
\end{equation}

\begin{figure*}[t]
\begin{center}
  \includegraphics*[width=4.25cm]{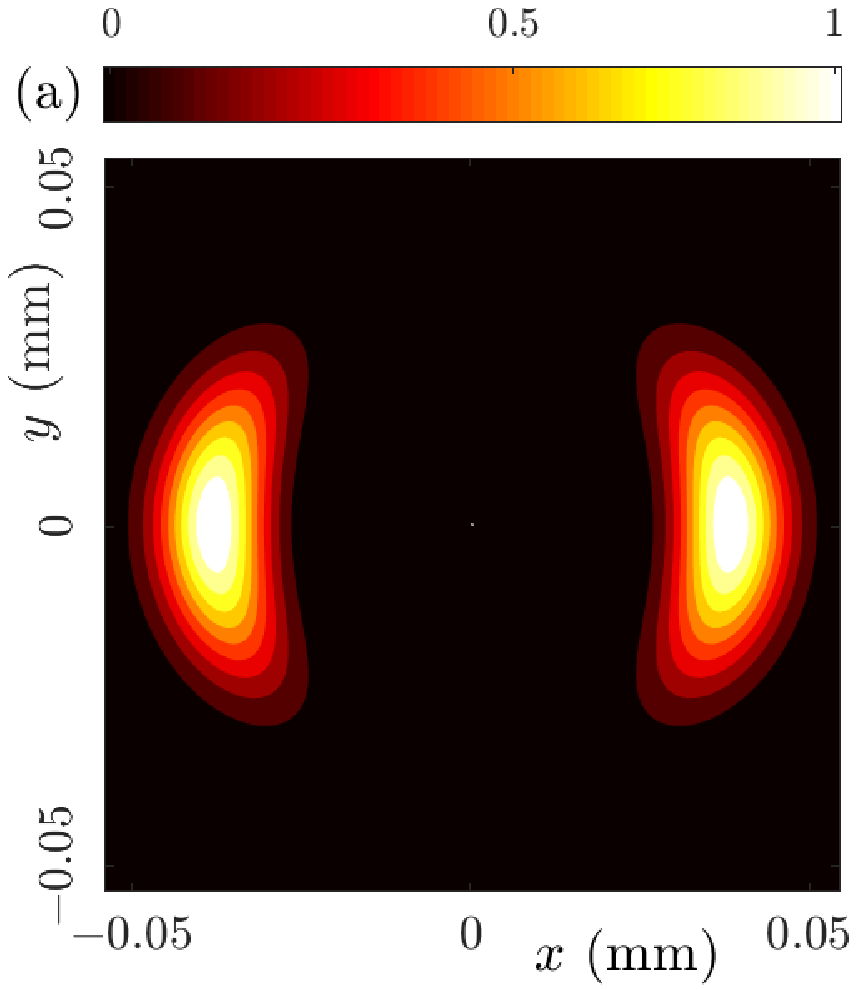}\includegraphics*[width=4.25cm]{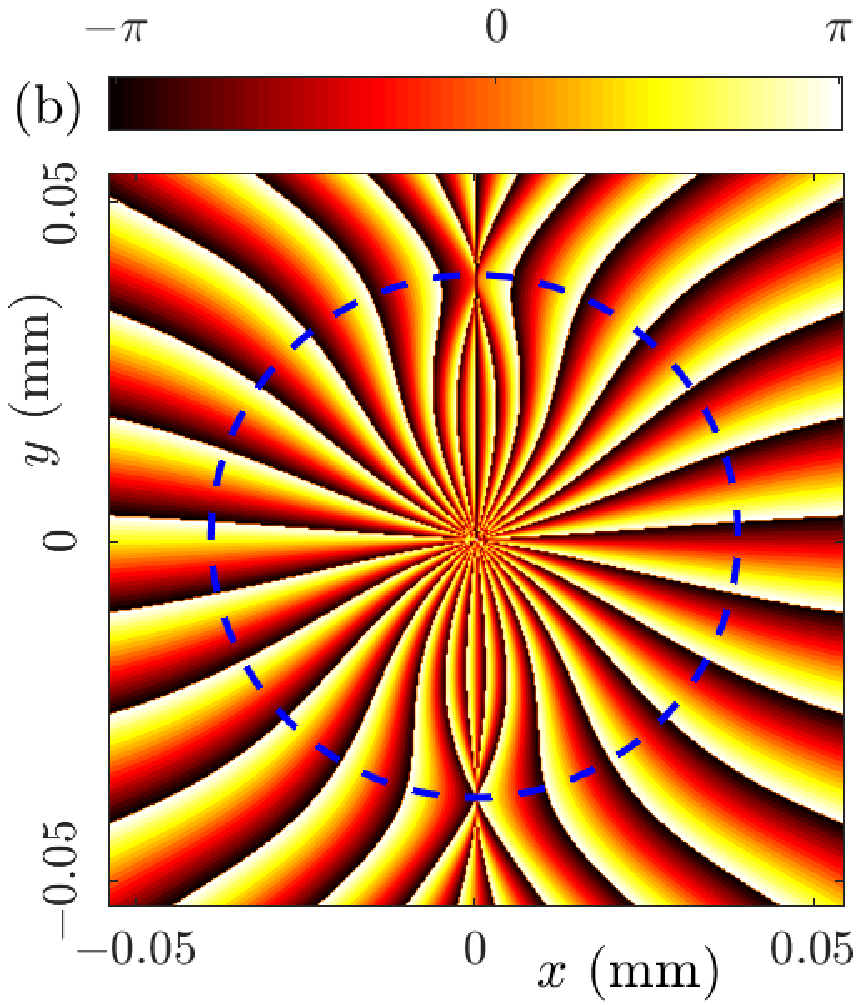}
  \includegraphics*[width=8.8cm]{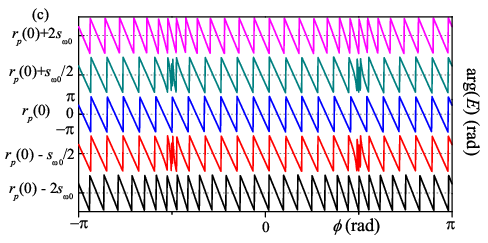}
\end{center}
\caption{\label{Fig3} (a) Amplitude $|E|$ and (b) phase $\mbox{arg}E$ of the helical pulse in Fig. \ref{Fig2} with $\delta l=1$ and $l_0=23$ at $z=0$ and $t'=0$. At the radius $r_p(0)$ where the intensity is maximum (dashed circle) the phase of the helical pulse is that of a vortex of the mean charge $l_0$. (c) Azimuthal variation of the phase $\mbox{arg}E$ at the indicated radii as a result of the topological charge dispersion.}
\end{figure*}

Under the above conditions, the sum in Eq. (\ref{AZ}) with the pulsed vortices in Eq. (\ref{FIXED}) (with $\omega_0$ replaced with $\omega_j$ and $l$ replaced with $l_j$) can be expressed, after straightforward algebra using Eqs. (\ref{LJ}) and (\ref{COND}), as
\begin{eqnarray}\label{HELICAL}
E &=& \frac{e^{-i(|l_0|+1)\psi(z)}e^{il_0\phi}}{\sqrt{1+\left(\frac{z}{z_R}\right)^2}}\left[\frac{r}{r_p(z)}\right]^{|l_0|}   A\left(t_c\right)e^{-i\omega_0t_c} \nonumber\\
&\times& a\left[t_c - \frac{l_0}{\omega_0}\phi + \frac{|l_0|}{\omega_0}\left(\psi(z)+ i\ln \frac{r}{r_p(z)}\right)\right],
\end{eqnarray}
where
\begin{equation}\label{CT}
t_c=t'- \frac{r^2}{2cq(z)}+i\frac{|l_0|}{2\omega_0}\,,
\end{equation}
and
\begin{equation}\label{AT}
a(t)=\sum_{|j|<\omega_0/\delta\omega} \tilde a_j e^{-i\delta\omega j t}\,.
\end{equation}
Condition $|j|<\omega_0/\delta\omega$ in Eq. (\ref{AT}) limits the sum to positive frequencies $\omega_j$.
Equation (\ref{HELICAL}) synthesizes the main result of this paper, and represents a helical pulse whose spatiotemporal structure under physically relevant conditions and propagation properties are discussed below. Being Eq. (\ref{HELICAL}) a finite sum of regular and three-dimensional localized pulsed vortices, the helical pulse is also regular and localized. The apparent singularity of the logarithm at $r=0$ gives, on account of Eq. (\ref{AT}) and the first row in Eq. (\ref{HELICAL}), the regular factor $(r/r_p(z))^{|l_0|+ j|\delta l|}$.

For ulterior use, the real and imaginary parts of the space-dependent, complex time in Eq. (\ref{CT}) can explicitly be separated as
\begin{equation}
t_c= t^{\prime\prime} - i \frac{r^2}{\omega_0s_{\omega_0}^2(z)} + i \frac{|l_0|}{2\omega_0},
\end{equation}
where $t^{\prime\prime}=t'-r^2/2cR(z)$. The real quadratic term $r^2/2cR(z)$ represents a time delay for the whole helical pulse structure to reach the distance $z$ at a radius $r$ due to the spherical pulse fronts of radius $R(z)$ when the pulse is converging to or diverging from the waist, as for the fundamental pulsed Gaussian beam \cite{PORRAS1,PORRAS2}. The dependence of the imaginary part on $l_0$ reflects the coupling between the OAM and temporal degrees of freedom, as recently described \cite{PORRAS6,CONTI}.

If the phases of $\tilde a_j$ are approximately constant, $a(t)$ represents a train of pulses with repetition period $\delta t =2\pi/\delta j\delta\omega$, where $\delta j$ is the step in the index $j$, e. g., $\delta j=2$ in high harmonic generation experiments. The bandwidth,
\begin{equation}
\Delta \omega_a = 2\left[\frac{\sum_j |\tilde a_j|^2(\omega_j-\omega_0)^2}{\sum_j |\tilde a_j|^2}\right]^{1/2}\,,
\end{equation}
of the train of pulses $a(t)$ is larger, and the duration $\Delta t_a$ of each one smaller, as more frequencies are superposed. If at least a few frequencies $\omega_j$ are superposed, the sorting $\Delta\omega_A<\delta\omega<\Delta\omega_a$ of the different frequency scales, and the opposite sorting $\Delta t_A> \delta t>\Delta t_a$ of the temporal scales, are satisfied.

An useful example with $\delta j=2$ is
\begin{equation}\label{COS}
a(t)=\cos^n(\delta\omega t)\,
\end{equation}
with $n$ even and $n<\omega_0/\delta\omega$, corresponding in Eq. (\ref{HELICAL}) to the superposition of $n+1$ frequencies ($n/2$ above and $n/2$ below $\omega_0$) spaced $2\delta\omega$. It can be seen that the coefficients $\tilde a_j$ (which can be found elsewhere) form an approximate Gaussian distribution of bandwidth $\Delta \omega_a\simeq \sqrt{2n}\,\delta\omega$. Correspondingly, each pulse in the train approximates the Gaussian shape $e^{-t^2/\Delta t^2_a}$ of diminishing duration $\Delta t_a\simeq  2/\Delta\omega_a = \sqrt{2/n}/\delta\omega$ as $n$ increases. Another example, mimicking the plateau region of a high harmonic spectrum, is $n+1$ frequencies (also $n/2$ above and $n/2$ below $\omega_0$) spaced $2\delta\omega$ with equal amplitudes $\tilde a_j$ and with approximate frequency bandwidth $\Delta\omega_a\simeq (2/\sqrt{3})\delta\omega n$. For not small $n$, each pulse in the train acquires the approximate form $\mbox{sinc}(t/T_a)$ of decreasing duration $T_a=2\pi/(\sqrt{3}\Delta\omega_a)= \pi/n\delta\omega$ [the first zero of $\mbox{sinc}(t/T_a)$] as $n$ increases.

At the bright ring, $r_p(z)$, Eq. (\ref{HELICAL}) for the helical pulse simplifies to
\begin{eqnarray}\label{RING}
E &=& \frac{e^{-i(|l_0|+1)\psi(z)}e^{il_0\phi}}{\sqrt{1+\left(\frac{z}{z_R}\right)^2}} A(t^{\prime\prime}) e^{-i\omega_0 t^{\prime\prime}} \nonumber \\
 &\times& a\left[t^{\prime\prime} - \frac{l_0}{\omega_0}\phi + \frac{l_0}{\omega_0}\psi(z)\right]\,. \label{RING}
\end{eqnarray}
The pulse shape at $r_p(z)$ and at fixed azimuthal angle $\phi$ then consists of the train of pulses $a$ of the carrier frequency $\omega_0$, duration $\Delta t_a$ and repetition period $\delta t$, enveloped by $A$ of the longer duration $\Delta t_A$, as in the two examples in Figs. \ref{Fig2}(a). As a function $\phi$ at fixed time, the angular period is $\delta\phi=(\omega_0/|l_0|)\delta t = 2\pi/\delta j \delta l$, i. e., each transversal section displays $N=\delta j\delta l$ spots, as seen in Fig. \ref{Fig2}(b). All together, at given distance, e. g. $z=0$, the helical pulse in Eq. (\ref{HELICAL}) has $N=\delta j\delta l$ equally spaced spots of light placed about the radius $r_p(0)$ that rotate in time counterclockwise (for $l_0>0$) or clockwise (for $l_0<0$) at the angular velocity $\Omega= \omega_0/l_0$ and that appear and disappear in the lapse of time $2\Delta t_A$. Plotted in the transversal and temporal dimensions, as in Fig. \ref{Fig2} (c) and (d), the whole structure is constituted by $N$ intertwined helices of pitch $2\pi/|\Omega|=2\pi |l_0|/\omega_0$ of finite duration $2\Delta t_A$. At any other distance $z$, as in Figs. \ref{Fig2} (e) and (f), the intertwined helices are expanded radially to $r_p(z)$, attenuated by diffraction, and rotated as a whole by the angle $\psi(z)$ as an effect of Gouy's phase shift. It is interesting to note that the pitch of intertwined attosecond {\it intensity} helices is the same as the pitch of intertwined helicoidal {\it phase front} of the fundamental pulse of frequency $\delta\omega$ and charge $\delta l$.

\begin{figure}[t]
\begin{center}
  \includegraphics*[width=4.25cm]{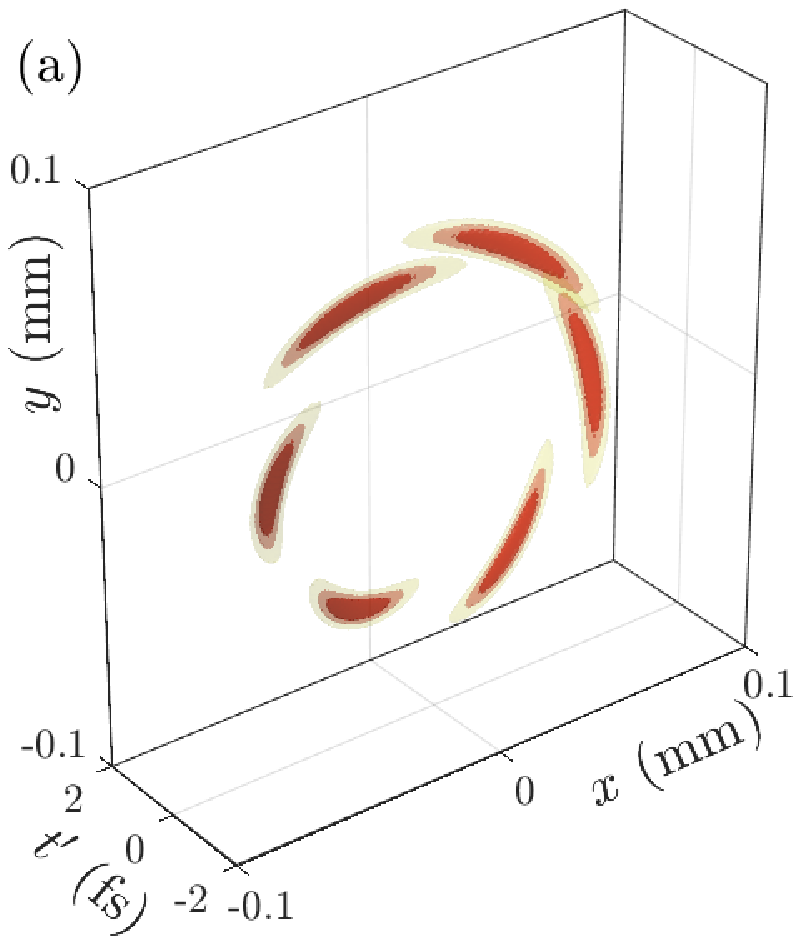}\includegraphics*[width=4.25cm]{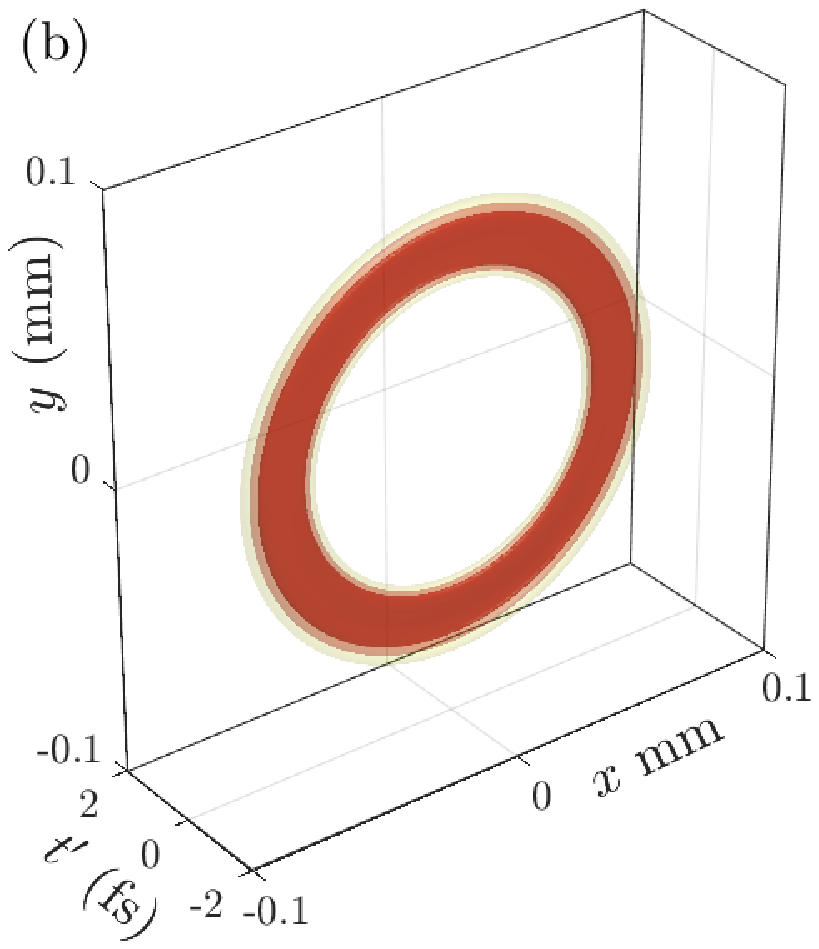}
  \includegraphics*[width=8cm]{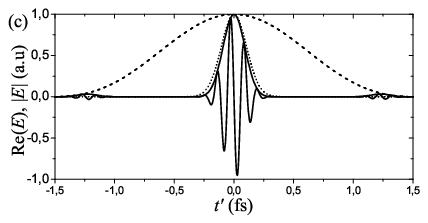}
  \end{center}
\caption{\label{Fig4} (a) Spatiotemporal structure of the intensity of the helical pulse with $\delta\omega=2.4166$ rad/fs$^{-1}$, $\omega_0=23\delta\omega=55.582$ rad/fs$^{-1}$, $\delta l=3$, $l_0=23\delta l=69$, $A(t)e^{-i\omega_0 t}=\mbox{sinc}^2(t/T_A)e^{-(i\omega_0t +i\pi/2)}$ with $T_A=1.5$ fs, $z_R=10$ mm, and $a(t)=\cos^n(\delta \omega t)$ with $n=20$ satisfying condition (\ref{CONDITION}), $15.33<n<23$. The three surfaces have intensities 0.2, 0.4 and 0.6 times the peak intensity. (b) For comparison, spatiotemporal structure of the shortest Gaussian-like pulse of the same carrier frequency carrying a vortex of topological charge $l_0=69$ without topological charge dispersion. (c) Train of pulses (real field $\mbox{Re}(E)$ and envelope $|E|$) at $\phi=0$ and $z=0$ (solid curves), enveloped by $|A|^2$ (dashed curve). Due to the sufficiently short duration of $A$, the train of pulses is an almost isolated attosecond pulse of duration $131$ as. The duration of the shortest Gaussian-like pulse carrying the charge $l_0=69$ without dispersion (dotted curve) is $149$ as.}
\end{figure}

\section{Narrowing attosecond helical pulses via topological charge dispersion}

\begin{figure}[t]
\begin{center}
  \includegraphics*[width=4.5cm]{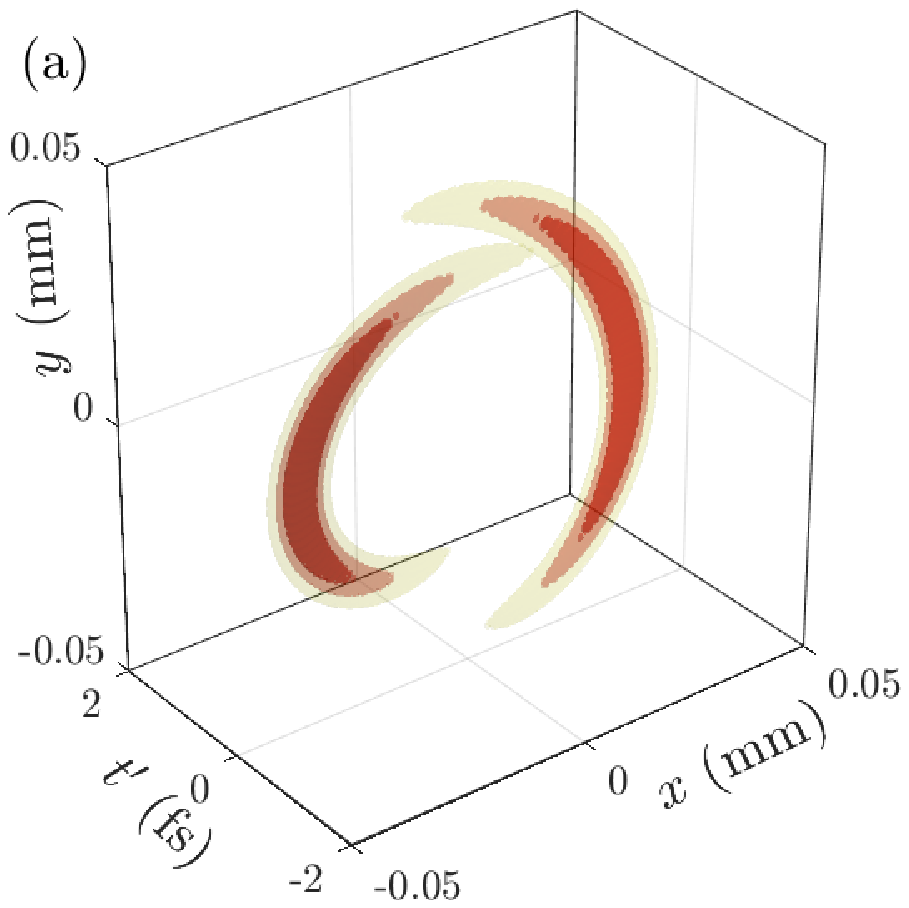}\includegraphics*[width=4cm]{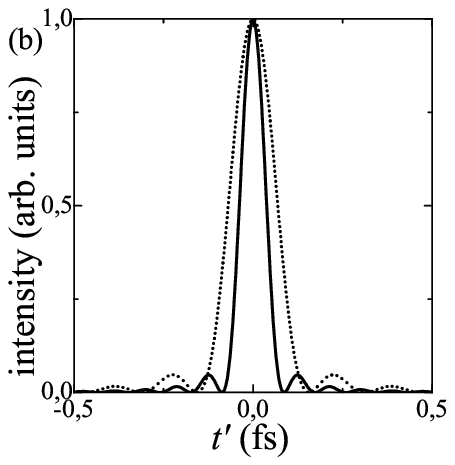}
  \end{center}
\caption{\label{Fig5} (a) Spatiotemporal structure of the intensity of the helical pulse with $\delta\omega=2.4166$ rad/fs$^{-1}$, $\omega_0=23\delta\omega=55.582$ rad/fs$^{-1}$, $\delta l=1$, $l_0=23\delta l=23$, $A(t)e^{-i\omega_0 t}=\mbox{sinc}^2(t/T_A)e^{-(i\omega_0t +i\pi/2)}$ with $T_A=1.5$ fs, $z_R=10$ mm, and $a(t)$ is made of $n=14$ frequencies with equal amplitudes about $\omega_0$ satisfying condition (\ref{CONDITION2}), $8.3<n<23$. The three surfaces have intensities 0.2, 0.4 and 0.6 times the peak intensity. (b) Intensity at $\phi=0$ and $z=0$ (solid curve) compared to the intensity of the shortest pulsed vortex of the same topological charge without dispersion (dotted curve).}
\end{figure}

As stated above, the bandwidth $\Delta \omega$ at the bright ring $r_p(z)$ of a pulsed vortex with a well-defined topological charge $l_0$ always satisfies inequality $\Delta\omega^2< 4\omega_0^2/|l_0|$, and this upper bound imposes a lower bound to the pulse duration \cite{PORRAS5}. A helical pulse presents however a dispersion in the topological charge about $l_0$ given by
\begin{equation}\label{LDISP}
\Delta l_a = 2\left[\frac{\sum_j |\tilde a_j|^2 (l_j-l_0)^2}{\sum_j |\tilde a_j|^2}\right]^{1/2} = \frac{|l_0|}{\omega_0}\Delta\omega_a\,,
\end{equation}
where the last relation follows from Eq. (\ref{COND}), and this dispersion makes the transverse phase pattern quite more complicated than the simple linear azimuthal variation $l_0\phi$, as illustrated in the example of Fig. \ref{Fig3}. Still, at the ring of radius $r_p(z)$ of maximum intensity [dashed circle in Fig. \ref{Fig3}(b)] the azimuthal variation continues to present the linear variation $l_0\phi$ [blue curve in Fig. \ref{Fig3}(c)] of a well-defined topological charge $l_0$.

On the other hand, if sufficiently high number of frequencies are superposed, the bandwidth $\Delta \omega$ and duration $\Delta t$ of the pulse in Eq. (\ref{RING}) at $r_p(z)$ are substantially the same as those of $a(t)$, i. e., $\Delta \omega\simeq \Delta\omega_a$ and $\Delta t\simeq \Delta t_a$. Interestingly, $\Delta \omega_a$ depends on $\delta \omega$ and the number of superposed frequencies, but is independent of $l_0$, which opens up the possibility to synthesize helical pulses verifying the opposite inequality $\Delta \omega\simeq \Delta\omega^2_a > 4\omega_0^2/|l_0|$ at $r_p(z)$, and thus to beat the lower bound to the pulse duration of the dispersion-free pulsed vortex, while retaining its azimuthal linear variation $l_0\phi$ at $r_p(z)$. From Eq. (\ref{LDISP}) with $\Delta\omega^2_a > 4\omega_0^2/|l_0|$, the required topological charge dispersion is
\begin{equation}
\Delta l_a>2\sqrt{|l_0|}.
\end{equation}

In the model with $a(t)=\cos^n(\delta\omega t)$, $n<\omega_0/\delta \omega$, inequality $\Delta\omega_a^2>4\omega_0^2/|l_0|$ with $\Delta\omega_a^2 = 2n\delta\omega^2$ leads to
\begin{equation}\label{CONDITION}
\frac{2}{|\delta l|}\frac{\omega_0}{\delta\omega} < n < \frac{\omega_0}{\delta\omega}
\end{equation}
for the number of frequencies about the carrier frequency, a condition that requires $|\delta l|>2$ to be satisfied. Since each pulse in the train has an approximate Gaussian shape of duration $\Delta t_a \simeq 2/\Delta \omega_a$, the lower bound $\Delta t_a > \sqrt{|l_0|}/\omega_0$ to the duration of a Gaussian-shaped, dispersion-free pulsed vortex turns into $\Delta t_a < \sqrt{|l_0|}/\omega_0$ for helical pulse if $n$ satisfies (\ref{CONDITION}). In the example of Fig. \ref{Fig4} satisfying (\ref{CONDITION}), the helical pulse at each azimuthal angle and propagation distance is a train of pulses of duration $\Delta t_a =131$ as [Fig \ref{Fig4}(a) and solid curves in Fig. \ref{Fig4}(c)], while the minimum duration of a pulse of the same carrier frequency carrying a vortex of charge $l_0=69$ without dispersion is $\Delta t_a =149$ as [Fig. \ref{Fig4}(b) and dotted curve in Fig. \ref{Fig4}(c)]. In addition, the envelope $A$ is taken sufficiently short [dashed curve in Fig. \ref{Fig4}(c)] so that the train of attosecond pulses reduces to an almost isolated attosecond pulse.

In the model with $n$ constant amplitudes $\tilde a_j$ about $\omega_0$ spaced $2\delta\omega$ and with $n<\omega_0/\delta\omega_0$, pulse shortening is more pronounced and is not restricted to $|\delta l|>2$. Condition $\Delta\omega_a>4\omega_0^2/|l_0|$ with $\Delta\omega_a\simeq (2/\sqrt{3})\delta\omega n$ leads now to
\begin{equation}\label{CONDITION2}
\sqrt{\frac{3}{|l_0|}}\frac{\omega_0}{\delta\omega} < n < \frac{\omega_0}{\delta\omega}\, .
\end{equation}
Since each pulse in the train has the approximate shape $\mbox{sinc}(t/T_a)$ with $T_a=2\pi/(\sqrt{3}\Delta\omega_a)= \pi/n\delta\omega$, the lower bound $T_a> \sqrt{\pi/3}\, \sqrt{|l_0|}/\omega_0$ to the duration without topological charge dispersion turns into $T_a< \sqrt{\pi/3}\, \sqrt{|l_0|}/\omega_0$ if $n$ satisfies (\ref{CONDITION2}). The helical pulse of Fig. \ref{Fig5}(a) with $l_0=23$ and $\delta l =1$ satisfies condition (\ref{CONDITION2}). At each particular azimuthal angle an isolated attosecond pulse of duration $T_a=93$ as appears [solid curve in Fig. \ref{Fig5}(a)], while the minimum duration of a sinc pulse of the same carrier frequency and topological charge without dispersion is $T_a= 157$ as (dotted curve). 

In the two examples above the envelope $A$ is taken with duration $\Delta t_A$ diminishing down to $\delta t$, or $\Delta \omega_A$ increasing up to $\delta \omega$, for the attosecond pulse to be isolated. In an experiment, a visible or near infrared driving pulse of envelope $A$ and charge $\delta l$ necessarily satisfies $\Delta\omega_A^2 < 4\delta\omega^2/|\delta l|$, which with $\Delta \omega_A\sim \delta\omega$ yields the limit $|\delta l| < 4$ to the topological charge of the fundamental pulse so that the attosecond pulse may be isolated.

\section{Conclusion}

In conclusion, we have provided a closed-form analytical expression that describes the attosecond helical pulses generated in recent experiments. Equation (\ref{HELICAL}) allows to understand the propagation of these attosecond helices of radiation and establishes a starting point for theoretical analyses of propagation in matter and other phenomena of interaction with matter.

In our analysis, a focusing geometry in which the fundamental pulse has a Rayleigh range or focal depth independent of frequency (and hence a frequency-dependent waist width) is assumed. A common Rayleigh range for all superposed harmonics arises naturally as the condition for their bright rings to overlap with the bright ring of the fundamental pulse. Attosecond helices of light with substantially the same properties have been described to be generated using other focusing geometries, e. g., with frequency-independent waist width in Ref. \cite{HERNANDEZ}, and are expected to arise with more sophisticated focusing configurations \cite{TOSA} since in all cases the crucial property of the attosecond helical pulses (the linear variation of topological charge with harmonic frequency) is imposed by conservation of angular momentum and is independent of the focusing geometry.

We have also shown that the helical pulses can transport vortices of  arbitrarily high mean topological charge and have at the same time arbitrarily short duration by virtue of the inherent topological charge dispersion. Possible generalizations of Eq. (\ref {HELICAL}), such as pulses carrying vortices with fractional topological charge \cite{TURPIN} or self-torque \cite{REGO2}, are currently under investigation.

The author acknowledges support from Projects of the Spanish Ministerio de Econom\'{\i}a y Competitividad No. MTM2015-63914-P, and No. FIS2017-87360-P.

\end{document}